# Applying voxel-based analysis to oropharyngeal cancer proton therapy patients: a correlation study on radiation-induced acute dysphagia

Qianxia Wang[1+], Alexander Stanforth[1], William Andrew LePain[1], Edgar Gelover[1], Haijian Chen[1], Mingyao Zhu[1], Katja M. Langen, Mark McDonald[1], James Edward Bates[1]*, Stella Flampouri[1]*

[1] Department of Radiation Oncology, Winship Cancer Institute of Emory University, Atlanta, GA 30322, United States of America

Emory Proton Therapy Center, Atlanta, GA 30308, United States of America

**\* Corresponding authors:** Stella Flampouri; e-mail, stella.flampouri@emory.edu. James Edward Bates; e-mail, `james.edward.bates@emory.edu.

**+ Statistical analysis author:** Qianxia Wang; e-mail, jcwlxj@gmail.com.

**Background:** Voxel-based analysis (VBA) is an analytic approach to evaluate correlations between local dose and the development of different toxicities. DVHs are used for toxicity prediction as well. Compared with DVH, no contours are required for VBA technique and results tell specific voxels that may be related to the toxicity instead of the whole contoured area. The VBA has been used on different cancer sites and for different toxicities. Most of these studies included patients treated with photon, all published studies were based on planned dose and VBA tools used were developed in house. In our study, patient cohort were treated with proton, our VBA tool was developed based on RayStation and doses fed to the VBA tool were delivered doses with constant and two variable RBE models.

**Purpose:** To develop a voxel-based analysis (VBA) tool with RayStation® scripting to evaluate correlations between spatial dosimetry and acute dysphagia in oropharyngeal cancer patients treated with chemoradiotherapy.

**Methods:** Forty-two oropharyngeal cancer patients treated with intensity-modulated proton therapy were included. All patients received bilateral neck irradiation and had no prior HN surgery or radiation. Clinician-graded acute dysphagia was recorded using CTCAE v5.0. Each patient's dose was estimated as a weighted sum of doses based on the planning CT and subsequent quality assurance CTs acquired during treatment for dosimetric effect evaluation related to anatomic changes. A constant and two variable RBE models provided estimations of biological doses. Each cumulative dose was mapped into a synthetic CT (common coordinate system) that represents this patient cohort. The cohort was split into toxicity and non-toxicity groups according to their toxicity grade. Permutation tests with maximum statistic strategy were performed on mapped doses to find locations with significant dose differences between the two groups. A loosened p-value threshold (0.1) was selected for sensitive cluster identification to conservatively include more area in identified clusters. The non-dosimetric variables (age and sex) on toxicity were studied with the Mann-Whitney U test.

**Results:** No sensitive cluster was detected with acute dysphagia grade$\geq$3 grouping criterion. For grade$\geq$2, one or two clusters depending on RBE models (one within the oral cavity and the other at the anterior-inferior part of the larynx) were identified. For non-dosimetric variables, only marginal correlation (p=0.07) between age and acute dysphagia was detected with grade$\geq$2.

**Conclusion:** In the patient cohort evaluated in this study, our VBA tool highlighted oral cavity and larynx subregions that may be related to the acute dysphagia development during definitive radiation for oropharyngeal cancer. Optimization of treatment plans to reduce dose to these areas may allow for a reduction in acute dysphagia in future patients.

# 1. INTRODUCTION

Modern radiation treatment techniques, such as IMRT, VMAT, and proton radiotherapy, utilize inverse planning, deliver conformal radiation dose to meet coverage goals for target volumes and minimize radiation to meet dose constraints to organs at risk (OARs). These OARs are typically anatomically defined, and dose constraints are based primarily on historical data and meta-analysis[1]. The therapeutic ratio for treatment is improved when treatment plans maintain target coverage while minimizing radiation dose to OARs. Despite improved treatment planning, patients experience acute and late radiation side effects from head and neck radiotherapy. While some radiation toxicities might be unavoidable given the prescribed doses and target positions, the frequency and intensity of some toxicities may be reduced by optimizing dose distributions to the most sensitive tissues. When using dose volume histogram (DVH) curves as a standard plan evaluation approach, sub- or multi-organ spatial dose correlations are potentially masked.

Voxel-based analysis[2–4] (VBA) is an analytic approach to evaluate correlations between local dose and the development of different toxicities. DVHs are used for toxicity prediction in some literatures[5–8] as well. However, no contours are required for VBA technique and results tell specific voxels that may be related to the toxicity instead of the whole contoured area from DVH. VBA comprises two parts: mapping the dose distribution of each patient to an anatomical common coordinate system (CCS) and statistically analyzing dose voxel-by-voxel for patients with and without toxicity.

VBA has been successfully applied in several disease sites, including head and neck cancer[9,10], prostate cancer[2,11,12], non-small cell lung cancer[13,14], Hodgkin lymphoma[15], and brain tumors[16], with corresponding dose areas correlated to different toxicities. Acute dysphagia was the toxicity investigated in this study.

Acute dysphagia is common and challenging to manage for patients undergoing definitive RT for oropharyngeal cancer. In RTOG 1016, approximately 35% of patients developed grade 3 or higher acute dysphagia, with approximately 60% needing a feeding tube placed by the end of treatment[17]. Acute dysphagia can lead to weight loss during treatment; increased weight loss during definitive therapy may be correlated with worse overall survival[18] and could be a precursor of late swallowing problems[19]. This underscores the importance of improving methods to reduce acute dysphagia during treatment.

To our knowledge, all published dosimetric correlation studies on radiation-induced toxicity using VBA were performed using different in-house tools. In the current study, we developed a VBA tool within the RayStation® scripting module, a standard commercial TPS system. Some core functions, such as dose calculation, image registrations and dose mapping, are called directly from the TPS which makes the tool development process much easier. Other functions required by the VBA technique, such as CCS selection, dose accumulation with different weights and statistics, were developed within RayStation scripting module. So the other potential advantage for this tool is that it is integrated into the TPS where CTs, doses and plans required by the VBA technique are stored, no extra software is needed, and clinic people are more familiar with operating TPS than other software. These advantages make it possible to widely disseminate programmed VBA techniques among different institutions in the future and simplify the inter-institutional data sharing and facilitates the VBA analysis for different beam field arrangements, more treatment sites, and various treatment modalities.

Toxicity-sensitive areas identified from the VBA analysis are strongly associated with doses received by the patient cohort. Therefore, the treatment radiation modality, the field arrangement, and the accuracy of the dose calculation largely determine the VBA results. Proton therapy attracts more and more interest due to the low entrance dose, no exit dose and higher biological effect compared to conventional photon therapy. To obtain a more accurate dose received by patients in this study, anatomy changes were considered in dose calculation by including the treatment planning CT (TPCT) and all other CTs collected during the treatment. For proton therapy, a constant RBE of 1.1 is clinically applied to conservatively estimate dose within the target. It ignores the influence of LET and other biological factors on RBE. Doses received by a patient were underestimated, especially at the distal end of proton beams.[20–23] To investigate the influence of different RBE model on VBA results, two variable and the constant RBE doses were applied for biological dose estimation.

Very few published VBA studies[13,24] discussed patients treated with proton therapy may due to the limited proton therapy availability. To our knowledge, no published studies applied VBA techniques on either variable RBE doses or cumulative delivered doses. In the current study, we applied our VBA tool developed within RayStation® system to a cohort of patients previously treated with IMPT for the base-of-tongue cancer at our center. Cumulative delivered doses with different RBE models were fed to the statistic tool to identify the spatial regions and dose levels that better correlated with the development of acute dysphagia.

## 2. MATERIALS AND METHODS

### 2.A. Patient cohort and treatment planning

Forty-two patients (Table 1) irradiated between 2019 – 2023 for intact and treatment-naïve disease (no prior surgery, no prior head and neck radiotherapy) and receiving bilateral neck radiotherapy with concurrent chemotherapy were included in this study. The dose and toxicity studies based on the 42 patients were conducted within the purview of a retrospective protocol approved by the Emory Internal Review Board.

Acute dysphagia symptoms of each patient were scored weekly during the treatment according to the Common Terminology Criteria for Adverse Events (CTCAE[25] v5.0). In this scale, the severity of acute dysphagia is described as Grade 1 for symptomatic patients but able to eat a regular diet; Grade 2 for altered eating/swallowing; Grade 3 for severely altered eating/swallowing, tube feeding, TPN, or hospitalization indicated; Grade 4 for life-threatening consequences, urgent intervention indicated and for Grade 5 for death. Among our cohort, 1 patient (2.4%) had no dysphagia symptoms during the whole treatment, 9 patients (21.4%) developed Grade 1 toxicity, 23 patients (54.8%) developed Grade 2 toxicity, and 9 patients (21.4%) developed Grade 3 acute dysphagia toxicity. No patients in this study developed Grade 4 or 5 dysphagia during treatment. And no patients received tube feeding during the treatment.

Toxicity was evaluated at two thresholds. We compared patients with grade 0-1 toxicity versus grade 2 and higher dysphagia toxicity (Dys2 cohort), and in a second analysis, evaluated those with grade 3 dysphagia (Dys3 cohort) against those with grade 0-2 toxicity, similar to prior studies[17, 18]. For a threshold of grade 2+ dysphagia, the Dys2 cohort included 32 patients (76%) versus 10 patients (24%) with grade 0-1 toxicity. For a threshold of grade 3 toxicity, the Dys3 cohort had 9 (21%) patients compared to 33 (79%) with grade 0-2 toxicity.

| Condition | Description | Patient # | Percentage |
|---|---|---|---|
| Age (years) | Range | 44-83 | |
| | Median | 70 | |
| Sex | Female | 6 | 14.3% |
| | Male | 36 | 86% |
| Dose to the primary site (Gy RBE) | 70/35 fx | 37 | 88% |
| | 69.96/33 fx | 4 | 10% |
| | 66/30 fx | 1 | 2% |
| Dose levels (SIB) | 3 levels | 22 | 52.4% |
| | 2 levels | 17 | 40.5% |
| | 1 level | 3 | 7.1% |
| Primary site laterality | Midline | 30 | 71.4% |
| | Right | 6 | 14.3% |
| | Left | 6 | 14.3% |
| Maximum acute dysphagia grade | 0 | 1 | 2.4% |
| | 1 | 9 | 21.4% |
| | 2 | 23 | 54.8% |
| | 3 | 9 | | 21.4% |
| Toxicity groups | | | |
| Grade 2+ Acute Dysphagia Toxicity (Dys2+ cohort) | (Grade 0, 1) | 10 | (4 midline, 5 right, 1 left) | 24% |
| | (Grade 2+) | 32 | (26 midline, 1 right, 5 left) | 76% |
| Grade 3 Acute Dysphagia Toxicity (Dys3 cohort) | (Grade 0, 1, 2) | 33 | (22 midline, 6 right, 5 left) | 79% |
| | (Grade 3+) | 9 | (8 midline, 1left) | 21% |
| Quality Assurance CTs obtained during treatment | 0-2 | 9 | | 21.4% |
| | 3 | 22 | | 52.4% |
| | 4-6 | 11 | | 26.2% |
| Patients with adaptive plans | 0 | 18 | | 43% |
| | 1 | 21 | | 50% |
| | 2 | 3 | | 7.1% |
| Extent stage | I | 8 | | 19% |
| | II | 15 | | 35.7% |
| | III | 11 | | 26.2% |
| | IV | 4 | | 9.5% |
| | No report | 4 | | 9.5% |
| T stage | T4 | 12 | | 28.6% |
| | T3 | 9 | | 21.4% |
| | T2 | 14 | | 33.3% |
| | T1 | 5 | | 11.9% |
| | T0 | 2 | | 4.8% |
| N stage | N3 | 2 | | 4.8% |
| | N2 | 13 | | 31.0% |
| | N1 | 21 | | 50.0% |
| | N0 | 6 | | 14.3% |
| HPV | Related | 10 | | 24% |
| | p16+ | 26 | | 61.9% |
| | p16- | 2 | | 4.8% |
| | no report | 4 | | 9.5% |

Table 1. Patient characteristics (SIB: Simultaneous Integrated Boost)

At our center, the standard practice of the head and neck radiation oncology group is to treat levels Ib, II-IVab, and Vab along with the retropharyngeal and retrostyloid spaces in the node-positive hemineck and to treat levels II-IVa in the node-negative hemineck. Minor deviations may have occurred on a case-by-case basis due to other clinical factors at the discretion of the treating radiation oncologist. The GTV for base of tongue patients included in this study is defined based on CT. MRI or/and PET/CT were also used for GTV delineation in some of these patients. The primary CTV is usually contoured as the GTV with a margin of 0.2 - 0.7 cm for the suspectable microscopic disease.

Bilateral head and neck cases are treated using a standardized five-field beam arrangement, which includes two anterior obliques, two posterior obliques, and an anterior-posterior (AP) field. The posterior obliques do not treat through the shoulders, and the contribution of the AP beam is kept 1 cm below the chin. While beam doses are modulated, at least two beams irradiate each part of the target. Plans are generated on the TPCT with a commercial TPS (RayStation®, RaySearch Laboratories, Stockholm). A Monte Carlo dose calculation algorithm within the TPS was used for both optimization and final dose calculation with a dose grid of 3 mm and energy range of 70.0 - 206.40 MeV. CTV robust optimization is used with 3 mm and 3.5% setup and range uncertainties, respectively. Plan quality was evaluated based on the institutional dose goals for OARs and 14 error scenarios (combined range and setup) for targets.

For verification purposes, we obtained quality assurance CT (QACT) throughout treatment. Typically, these are acquired every 10 fractions; however, the acquisition frequency may vary at the discretion of the treatment physician, depending on various clinical factors (for example, weight loss, anatomic change identified on an exam, or on cone-beam CT). In this cohort, patients had a median of 3 (range 0-6) QACT scans, with 97.6% having at least one QACT scan.

## 2. B. Dose accumulation

Each patient's total delivered dose was calculated based on the patient's TPCT and QACTs and accumulated with different weights to account for anatomic changes during the treatment course. The weights for different CTs were determined by the number of treatments delivered close to the CT acquisition date (Appendix A). Doses based on TPCT and all QACTs were calculated by calling the Monte Carlo Algorithm in RayStation® from the scripting module. After mapping all QACT based doses to the TPCT, accumulating doses with different weights were performed by scripting developed in RayStation®.

The constant RBE of 1.1 doses was calculated within the regular clinical RayStation® version, which is unable for the variable RBE dose calculation. So variable RBE doses were estimated within the RayStation® 11B (research version). The input data included variable RBE model type, reference dose per fraction and ($\alpha_x$, $\beta_x$) values for normal tissue and targets. In this study, we investigated two phenomenological RBE models: Wedenberg[27] and McNamara[28], and a reference fraction size of 2 Gy. The radiosensitivity parameters used for the calculation were $\alpha_x = 0.0499$ Gy$^{-1}$, and $\beta_x = 0.0238$ Gy$^{-2}$ [29] for normal tissues and $\alpha_x = 0.22$ Gy$^{-1}$, $\beta_x = 0.0272$ Gy$^{-2}$ [30] for targets.

Since the calculation of variable RBE on QACTs cannot be done directly on the platform used in this study, the original plan parameters (i.e., spot position, MU/spot, energy layers, etc.) were transferred to the QACT datasets through Python scripting to generate an updated dose distribution.

**2.C. Common coordinate system (CCS) selection, image registration and dose mapping**

A method was developed to search for an anatomical CCS among the patient cohort. The process was based on Dice score calculations for 6 critical structures between each pair of patients before the deformable registration (Appendix B). After the selection of a CCS (Figure 1 (a)), deformable registrations were performed between the CCS and each patient to (1) verify Dice scores of these deformed critical structures and (2) map doses from each patient's TPCT to the CCS.

Image registrations were performed by calling the rigid registration and hybrid DIR from RayStation in scripting module. The rigid registration of RayStation is based on points of interest. The hybrid DIR uses anatomically constrained deformation algorithm[31,32], which formulates the registration problem as a non-linear optimization problem and is GPU accelerated.

Each accumulated dose on the patient's TPCT were mapped to the CCS for statistical analysis across patients through the function in RayStaiton called by scripting. The quality of dose mapping was determined by the similarity between the individual patient and the CCS, demonstrated by high Dice scores after deformable registration, visual inspection of deformed CT and dose for each patient and the clear and shaped edges for the average deformed CT of all patients (Figure 1 (b)).

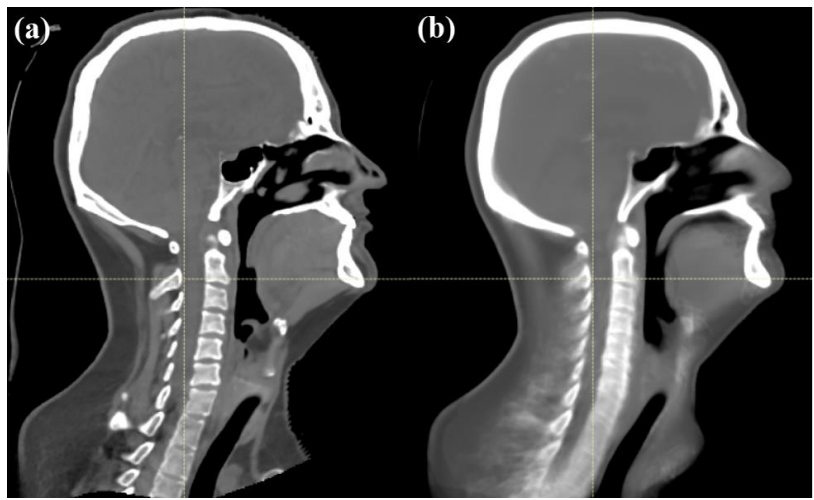

*Figure 1. The synthesized CCS (a) and the average CT of all patients after deformable registration (b).*

## 2.D. Statistical analysis

A multiple comparisons permutation test[33] was performed to identify areas with significant dose differences between groups with and without the threshold level of toxicity. In this permutation test, the statistics performed within each voxel was the difference in mean doses between the two groups normalized to the standard deviation of mean dose differences of this voxel for 10,000 permutations. The maximum statistical values over all voxels for each permutation were saved as expected values. The p-value for one voxel was the proportion of expected statistic values larger than the observed value. This statistical test can be carried out in the RayStation® scripting environment.

Even though we had a relatively even distribution of right-sided, left-sided, and midline tumors (Table 1), the dose distributions were not evenly distributed symmetrically due to the extent of disease, nodal distributions, and other anatomic factors. To generate symmetrical data, we mirrored each patient's TPCT and dose distribution and included them in our analysis[10].

We defined a pre-specified significant threshold for a voxel of p ≤ 0.1, a relatively loosened statistical threshold to compensate for the extremely strict p-value evaluation scale (maximum statistical values over all voxels for each permutation). We then contoured voxels with p ≤ 0.1 except those corresponding to air, bone, or subcutaneous fat, which we do not believe impact clinical dysphagia outcomes. For each voxel within the contour, the mean dose of each group (those with and without the threshold level of toxicity) was recorded. The Mann-Whitney U[34] test was used to compare group voxel mean doses for each area.

The Mann-Whitney U[34] test was also applied to find a correlation between the acute dysphagia and two non-dosimetric variables, sex and age, with criteria of Dys2 and Dys3.

The flowchart summarizes the overall workflow of the VBA tool for toxicity analysis (Figure 2).

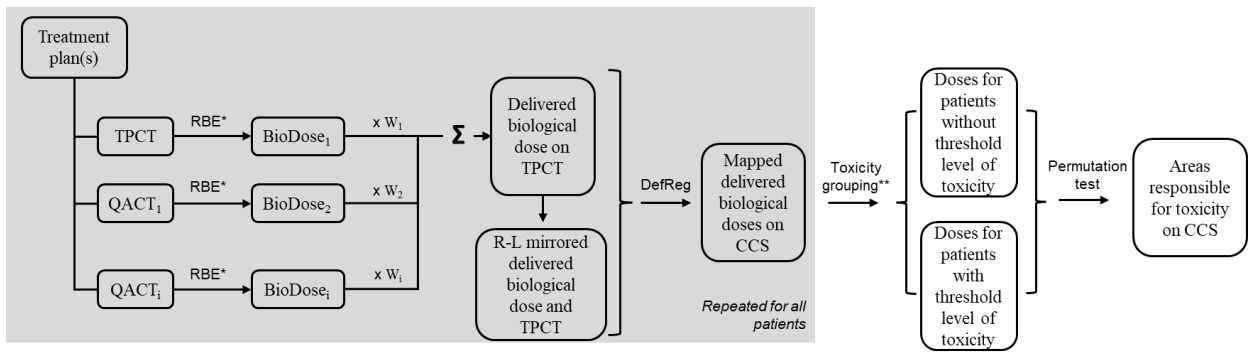

*Figure 2. Flowchart of the VBA tool for dysphagia toxicity analysis. (RBE\*: the constant RBE of 1.1, Wedenberg and McNamara models; Toxicity grouping\*\*: toxicity group with threshold levels of grade ≥2 and grade ≥3.).*

**3. RESULTS**

The Dice score averaged over the 6 critical structures of the selected CCS compared to patients in the cohort before the deformable registration ranged from 0.66 to 0.79, with a median value of 0.73 and a mean value of 0.72. The Dice score approached 1 after the deformable registration.

Three patients were picked up to verify that the accumulated dose calculated by scripting agreed with that from manual operation in RayStation (calculating weights by hand, inputting them into operation window and running the calculation). After mapping each accumulated dose to the CCS, we visually inspected deformed doses for about 50% of patients. All of them are smooth and consistent.

In this study, an analysis with two toxicity thresholds was performed. One threshold was grade 3, in which patients with grade 3 toxicity were assigned to the toxicity group, and those with grades 0-2 were assigned to the non-toxicity group (Dys3). No clusters were identified to be toxicity-related areas with this threshold. The other threshold was grade 2, in which patients with grade 0-1 acute dysphagia were compared with those experiencing grade 2-3 toxicity (Dys2). With this threshold, one or two clusters depending on RBE models were identified with significant dose differences between the two groups.

Using the threshold of grade 2+ toxicity, two clusters (the oral cluster and the larynx cluster) were identified with significant dose differences between groups for the constant RBE model. Only one cluster was identified in the oral cavity for the two variable RBE models (Wedenberg and McNamara). The mean dose, mean dose difference, and identified clusters for different RBE models are displayed in Figure 3. Cluster volumes for different RBE models, cluster overlaps between each pair of RBE models, and dose analysis within clusters are listed in Table 2.

The identified oral clusters had a volume of 9.7 cc (6.0% of oral cavity) for the constant RBE, 9.6 cc (6.0% of oral cavity) for the Wedenberg model, and 10.0 cc (6.2% of oral cavity) for the McNamara model. The oral clusters corresponding to different RBE models were at similar location and volume differences between each pair of RBE models were less than 5.0%. The oral cluster overlap between variable RBE models and the constant RBE was 85.0% for the Wedenberg model and 87.3% for the McNamara model. The cluster overlap between the two variable RBE models was 95.1%.

Within the oral cluster, the toxicity group had a mean dose (standard deviation) of 38.3 (11.3) Gy, 39.2 (10.3) Gy, and 40.4 (10.8) Gy for the constant RBE, Wedenberg, and McNamara models, respectively. The non-toxicity group has a mean dose (standard deviation) of 15.8 (12.0) Gy, 15.6 (10.7) Gy, and 16.8 (11.7) Gy for the constant RBE, Wedenberg, and McNamara models, respectively. The mean dose of variable RBE models was supposed to be slightly higher than that of the constant RBE models. However, it was not true for the non-toxicity group. The Wedenberg mean dose is 0.2 Gy lower than the constant RBE maybe due to the location and volume differences in their clusters.

Within the oral cluster, the mean dose difference between the toxicity and non-toxicity groups for each RBE model was as high as 22.5 Gy for the constant RBE model, 23.7 Gy for Wedenberg model, and 23.6 Gy for the McNamara model. The Mann-Whitney U test was applied to mean dose distributions between groups with and without grade 2+ toxicity for the identified oral clusters from different RBE models. All p-values were less than 0.0001, verifying the significant difference between the toxicity and non-toxicity groups within the oral clusters.

A second cluster with a volume of 1.0 cc (6.0% of larynx) around the larynx was identified using the constant RBE model. For the McNamara model, a tiny cluster with a less than 0.1 cc volume was found at the same location. For the Wedenberg model, no similar cluster was identified. For the larynx cluster identified with the constant RBE, the mean dose (standard deviation) was 14.6 (4.0) Gy for the toxicity group and 7.2 (2.2) Gy for the non-toxicity group. The dose difference was 7.4 Gy, which was also significant. Similarly, the Mann-Whitney U test was applied to compare mean doses of each voxel between the toxicity and non-toxicity groups, and a very low p-value (<0.0001) was obtained.

For the two non-dosimetric variables, age and sex, only a marginal correlation was found between age and acute dysphagia with the criterion of Dys2 (p=0.07). The average ages were 65 and 71 for the toxicity and non-toxicity group, respectively.

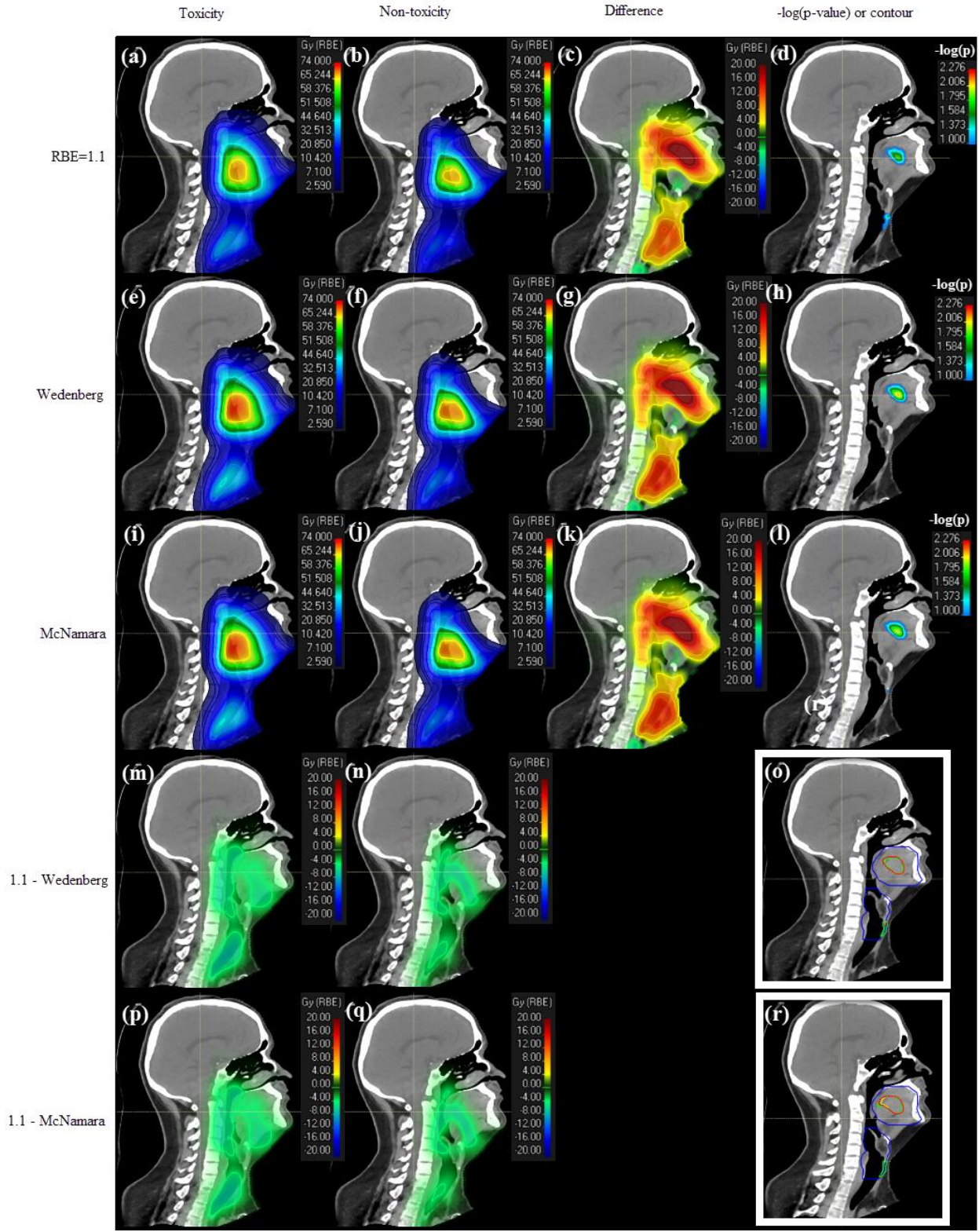

*Figure 3. Panels of the first 3 rows were for the constant RBE of 1.1, Wedenberg and McNamara: mean dose map in a sagittal plane (0.6 cm from the center) for those with grade 2+ toxicity (a, e, and i), those with grade 0-1 toxicity (b, f, and j), dose difference (toxicity group - non-toxicity group) (c, g, and k), -log(p) map (d, h, and l). Panels of the 4th (Wedenberg) and 5th (McNamara) rows were for mean dose difference between the constant RBE and variable RBE: (constant RBE – variable RBE) for those with (m and p) and without (n and q) grade 2+ toxicity. Panels o and r were for cluster contours (the constant RBE of 1.1 (green), the Wedenberg model (yellow), and the McNamara model (red)), oral cavity (blue) and larynx (blue) at two different sagittal planes.*

| Toxicity group definition | | grade≥2 | | |
|---|---|---|---|---|
| RBE models | | 1.1 | Wedenberg | McNamara |
| Oral cluster | Volume(cc) | 9.7 | 9.6 | 10.0 |
| | Cluster Overlap with RBE = 1.1 (%) | | 85.0 | 87.3 |
| | Cluster Overlap of the two variable RBEs (%) | | 95.1 | |
| | Mean dose and standard deviation for those with grade 2+ toxicity (Gy) | $38.3 \pm 11.3$ | $39.2 \pm 10.3$ | $40.4 \pm 10.8$ |
| | Mean dose and standard deviation for those with grade 0-1 toxicity (Gy) | $15.8 \pm 12.0$ | $15.6 \pm 10.7$ | $16.8 \pm 11.7$ |
| | p-value | <0.001 | <0.001 | <0.001 |
| larynx cluster | Volume (cc) | 1.0 | 0 | ~0 |
| | Mean dose and standard deviation for those with grade 2+ toxicity (Gy) | $14.6 \pm 4.0$ | | |
| | Mean dose and standard deviation for those with grade 0-1 toxicity(Gy) | $7.2 \pm 2.2$ | | |
| | p-value | <0.001 | | |

*Table 2. Cluster volume and overlap percentage between different RBE models, mean dose, and standard deviation of those with and without grade 2+ toxicity analyzed within each cluster for different RBE models.*

## 4. DISCUSSION

A VBA tool was developed within the RayStation® scripting module and applied to 42 base-of-tongue cancer patients treated with IMPT to analyze acute dysphagia toxicity. Three different RBE models were applied for delivered dose calculation, and two patient grouping criteria were investigated. One or two clusters depending on RBE models were identified as related to acute dysphagia toxicity.

The voxel-based analysis was performed on doses received by each patient. The accuracy of dose calculation plays a vital role in the statistical results. In our VBA tool, each patient's dose was the accumulation of doses based on the TPCT, and all QACTs acquired during the treatment with different weights. This dose accumulation accounted for the anatomical changes of patients during the whole treatment process, and it represents more accurately the dose received by a patient. The biological dose received for each patient was calculated with the constant RBE of 1.1 and two variable RBE doses.

Two criteria for grouping patients were studied. No toxicity-related cluster was identified with the criterion of grade 3 threshold. With the criterion of grade 2 as a threshold, two clusters were identified as related to toxicity based on delivered doses with a constant RBE of 1.1. One was centered in the posterior portion of the oral cavity. The other one included the bottom of the larynx and muscles anterior to the larynx. Minor salivary glands all over the mucosa of the larynx may be the contributing cause of acute dysphagia.

Using the two variable RBE models resulted in similar oral cavity regions as the constant RBE (with an overlap of ~ 85%). The Wedenburg and McNamara oral clusters were 1.5% smaller and 3.2% larger than the constant RBE model, respectively. The oral clusters identified by the two variable RBE models were almost identical (with an overlap of 95.1%, volume difference of 4.7%, and mean dose difference within clusters of 1.2 Gy). When using variable RBE models, the larynx cluster either disappeared (Wedenburg model) or became minuscule (~0.1cc, McNamara model) and ignored in the table due to too small volume.

In most voxels, variable RBE doses were higher than the constant RBE dose, especially at the end of the beam range, where the difference can be as high as 11 Gy.[35] As expected, generally, the variable RBE models had slightly higher mean doses than the constant RBE model within clusters (-0.2 Gy $\leq \Delta \overline{D} \leq$ 2.1 Gy). The identified oral clusters from doses were at a similar location and had comparable volumes for different RBE models. The identified larynx clusters differed greatly from the constant RBE and variable RBEs. This is because p-values used for cluster identification were calculated by comparing doses of two patient groups calculated within the same RBE model. The identified clusters from p-value maps can be similar or very different between the constant and variable RBEs. Therefore, it is imprudent to conclude that variable RBE models are of little or great significance for toxicity study using VBA.

From the results obtained in this study, we recommend applying both constant and variable RBE doses for future VBA toxicity studies for proton therapy if available. First, the contours of oral clusters from the constant and variable RBE models in this study are similar but not entirely superimposed (with about 15% volume difference). Second, the contours in this study are based on a p-value $\leq$ 0.1. Contours from different RBEs could diverge more with another p-value threshold because p-value maps differ per model. Thirdly, we obtained similar results in the oral cavity from different RBE models in our study, but results could be different for other toxicity, site analysis, or $\alpha_x$ and $\beta_x$ values. Lastly, the larynx cluster are completely different from different RBE doses.

The locations of the identified clusters could indicate that some subregions of organs, such as the oral cavity and larynx, are more sensitive to radiation. These findings can lead to the adoption of stricter clinical goals during planning optimization for the anatomy that overlaps with the clusters. Within the larynx cluster, the muscle anterior to the larynx was identified to correlate with dysphagia too.

Monti et al. in 2017[9] applied VBA on HN patients treated with photon therapy for acute dysphagia. In their study, two clusters around the cricopharyngeal muscle and cervical esophagus were identified with the criterion of Dys3. Neither of these clusters was identified in our study. Because patients in our study were treated with IMPT, which had a lower entrance dose and no exit dose, doses at these locations were much lower than their doses.

As far as we know, Monti's study in 2017 was the only publication for acute dysphagia analysis with VBA. The VBA technique was also used for the dosimetric correlation study on dysphagia in the work of Vasquez Osorio et al.[36] in 2023, where the toxicity was 1 year after the treatment instead of acute dysphagia. In their study, the patient grouping criterion was using MDADI scale with a threshold of 60, which was comparable to Dys2 used in our study. The inferior section of the brainstem was found to be strongly associated with the toxicity. This region was not identified in our study maybe due to (1) different toxicity studied (patients may be grouped differently according to acute and late dysphagia ); (2) different statistic test performed (p-value map changes with different statistic strategies) and (3) different treatment modalities (proton patients received very low dose at the area of brain stem compared to photon patients).

Another two studies applied the DVH technique to photon patients with HN cancer and found that acute dysphagia (with grade = 3) was related to the larynx DVH doses[6,7] and cricopharyngeal muscle with Dys2[5]. With Dys2, the larynx cluster identified in our study included the lower part of the larynx. However, no cluster at the cricopharyngeal muscle was identified due to the low dose to this muscle with proton plans.

The oral cluster (a subregion of the oral cavity) identified in our study was not shown in their analysis either. Our observed oral cluster is approximately at the edge of the typical primary site CTV and may be related to disease extension and larger range uncertainty in IMPT. However, for photon therapy, the overall increased oral cavity doses (due to high exit dose of each beam field) obscure a dose difference between those with and without dysphagia. Functions of the oral cavity, such as togue movement, dentition, buccal mucosa and saliva production play key role in swallowing. Radiation damage to this organ may cause dysphagia. [37]

Before closing the discussion, two limitations are worthy of note. One is the toxicity grade used for patient grouping. The grade selected for each patient is the worst scored weekly during the treatment course. Treatment toxicity grading is unavailable during the immediate weeks post-treatment when acute toxicity may worsen. So, the "acute dysphagia" discussed in this study only refers to toxicity during the treatment. And also, the baseline dysphagia was not deducted from toxicity grade because a different criterion system was used for toxicity evaluation. The other limitation is the number of patients involved in the final statistics. We obtained stable cluster positions with this cohort of patients. However, areas of clusters are still sensitive to the number of patients. Large cohorts are desired for more accurate cluster area prediction.

With accurate clusters identified from the VBA, we suggest combining oral clusters from the different RBE models and including this combined oral cluster and larynx cluster as extra OARs in planning optimization with stricter dose constraints. Dose constraints can be set as the first few percentiles of mean dose of the toxicity group within clusters. The CCS CT, cluster contours and new dose constraints can be shared with other institutions where base-of-tongue cancer patients are treated with the standardized five-field IMPT beam arrangement.

In summary, by applying VBA to HN proton plans, we identified specific anatomic regions and dose exposures that could be associated with acute dysphagia toxicity. The proposed VBA can be used to study different toxicities and cancer sites. Compared with other VBA studies, we used an estimation of the patient-received doses based on both TPCT and QACTs and with different RBE models for analysis. And compared with other VBA studies, the proposed VBA tool was developed based on RayStation®, a widely used TPS system in clinics. Most core functions of this VBA tool are called from the RayStation® and some functions were developed within RayStation® script, no more software is required. This VBA tool makes it easy to share processed patient doses from different institutions. Those doses can be calculated with the same dose engine, mapped with the same registration algorithm and to the same CCS.

For non-dosimetric variables, only a marginal correlation was found between age and acute dysphagia with the criterion of Dys2. Younger patients are inclined to suffer from grade 2-3 acute dysphagia.

However, this conclusion was based on the limited number of patients. To confirm this conclusion, further study is required.

**ACKNOWLEDGMENTS**

We thank Zachary Diamond, Roelf Slopsema, Shadab Momin, Mosa Pasha and Anees Dhabaan for helpful discussions and comments on this study. And we thank William Stokes and Soumon Rudra for their contribution to patient treatment.

**CONFLICT OF INTEREST**

We have no conflicts of interest to disclose.

**APPENDIX A**

The weights for different CTs were calculated in the following steps:

a) record fraction numbers from ARIA® (Varian Medical Systems, Inc.) on dates when QACTs were scanned, $I_1$, $I_2$, $I_3$..., where I was the fraction number and subscripts (1, 2, 3 …) were indices of QACTs sorted by date.

b) calculate the fraction number for the date between two successive QACTs ($QACT_j$ and $QACT_{j+1}$) with the equation

$$T_j = I_j + int(\frac{I_{j+1} - I_j}{2}); \tag{1}$$

c) calculate the weight for $QACT_j$ with the equation

$$w_j = T_j - T_{j-1}; \tag{2}$$

d) and assign only 1 as the weight for a TPCT because it was obtained 10 days before the first treatment.

## APPENDIX B

All TPCTs from the 42 patients were imported into a 'single patient instance' within RayStation®. Within this 'single patient instance', one TPCT was picked up as a reference TPCT ($TPCT_i$) and all the rest TPCTs ($TPCT_j$, $j=1, 2, 3... N$ and $j \neq i$) were rigidly registered to selected $TPCT_i$. Then the Dice scores of all critical structures (the spinal cord, left and right parotids, oral cavity, brain stem, and mandible bone) for each pair of patients ($P_i$ and $P_j$) were calculated and the averaged Dice score over these six structures was saved as the Dice score for this pair of patients ($\overline{D}_{ij}$). Finally, we repeated steps for all $j$s and the Dice score averaged over all $j$s was set as the Dice score for patient $i$ ($\overline{D}_i$).

The patient with the best Dice score was selected to prepare for the CCS. (There are two patients who have the same highest average Dice score ($\overline{D}_i$). The patient with fewer low pair Dice scores ($\overline{D}_{ij}$) was selected.) This patient could not be directly used to create the CCS due to gross disease in the head and neck. To avoid the distortions of gross disease in the CCS, CT images of a CNS patient with no abnormality or image artifacts in the head and neck region were deformed to match the selected head and neck patient's anatomy. The resulting images (the deformed CNS) were used as the CCS (Figure 1 (a)).

The average Dice score by comparing the CCS (the deformed CNS patient) to all patients is comparable to the selected patient with the best average Dice score.